# Efficient Subgraph Similarity Search on Large Probabilistic Graph Databases

Ye Yuan§  Guoren Wang§  Lei Chen†  Haixun Wang‡
§College of Information Science and Engineering, Northeastern University, China
†Hong Kong University of Science and Technology, Hong Kong, China
‡Microsoft Research Asia, Beijing, China
{yuanye,wanggr}@ise.neu.edu.cn, leichen@cse.ust.hk, haixunw@microsoft.com

## ABSTRACT

Many studies have been conducted on seeking the efficient solution for subgraph similarity search over certain (deterministic) graphs due to its wide application in many fields, including bioinformatics, social network analysis, and Resource Description Framework (RDF) data management. All these works assume that the underlying data are certain. However, in reality, graphs are often noisy and uncertain due to various factors, such as errors in data extraction, inconsistencies in data integration, and privacy preserving purposes. Therefore, in this paper, we study subgraph similarity search on large probabilistic graph databases. Different from previous works assuming that edges in an uncertain graph are independent of each other, we study the uncertain graphs where edges' occurrences are correlated. We formally prove that subgraph similarity search over probabilistic graphs is #P-complete, thus, we employ a *filter-and-verify* framework to speed up the search. In the *filtering* phase, we develop tight lower and upper bounds of *subgraph similarity probability* based on a probabilistic matrix index, PMI. PMI is composed of discriminative subgraph features associated with tight lower and upper bounds of *subgraph isomorphism probability*. Based on PMI, we can sort out a large number of probabilistic graphs and maximize the pruning capability. During the *verification* phase, we develop an efficient sampling algorithm to validate the remaining candidates. The efficiency of our proposed solutions has been verified through extensive experiments.

## 1. INTRODUCTION

Graphs have been used to model various data in a wide range of applications, such as bioinformatics, social network analysis, and RDF data management. Furthermore, in these real applications, due to noisy measurements, inference models, ambiguities of data integration, and privacy-preserving mechanisms, uncertainties are often introduced in the graph data. For example, in a protein-protein interaction (PPI) network, the pairwise interaction is derived from statistical models [5, 6, 20], and the STRING database (http://string-db.org) is such a public data source that contains PPIs with uncertain edges provided by statistical predications. In a social network, probabilities can be assigned to edges to model the degree of influence or trust between two social entities [2, 25, 14]. In a RDF graph, uncertainties/ inconsistencies are introduced in data integration where various data sources are integrated into RDF graphs [18, 24]. To model the uncertain graph data, a probabilistic graph model is introduced [27, 43, 21, 18, 24]. In this model, each edge is associated with an edge existence probability to quantify the likelihood that this edge exists in the graph, and edge probabilities are *independent* of each other. However, the proposed probabilistic graph model is invalid in many real scenarios. For example, for uncertain protein-protein interaction (PPI) networks, authors in [9, 28] first establish elementary interactions with probabilities between proteins, then use machine learning tools to predict other possible interactions based on the elementary links. The predictive results show that interactions are correlated, especially with high dependence of interactions at the same proteins. Given another example, in communication networks or road networks, an edge probability is used to quantify the reliability of link [8] or the degree of traffic jam [16]. Obviously, there are correlations for the routing paths in these networks [16], i.e., a busy traffic path often blocking traffics in nearby paths. Therefore, it is necessary for a probabilistic graph model to consider correlations existed among edges or nodes.

Clearly, it is unrealistic to model the joint distribution for the entire set of nodes in a large graph, i.e., road and social networks. Thus, in this paper, we introduce joint distributions for local nodes. For example, in graph 001 of Figure 1, we give a joint distribution to measure interactions (neighbor edges[1]) of the 3 nodes in a local neighborhood. The joint probability table (JPT) shows the joint distribution, and a probability in JPT (the second row) is given as $Pr(e_1 = 1, e_2 = 1, e_3 = 0) = 0.2$, where "1" denotes existence while "0" denotes nonexistence. For larger graphs, we have multiple joint distributions of nodes in small neighborhoods (in fact, these are marginal distributions). In real applications, these marginal distributions can be easily obtained. For example, authors in [16] use sampling methods to estimate a traffic joint probability of nearby roads, and point out that the traffic joint probability follows a multi-gaussian distribution. For PPI networks, authors in [9, 28] establish marginal distributions using a Bayesian prediction.

In this paper, we study subgraph similarity search over probabilistic graphs due to wide usage of subgraph similarity search in many application fields, such as answering SPARQL query (graph) in RDF graph data [18, 1], predicting complex biological interactions (graphs) [33, 9], and identifying vehicle routings (graphs) in road networks [8, 16]. In the following, we give the details about subgraph similarity search, our solutions and contributions.

---

[1]Neighbor edges are the edges that are incident to the same vertex or the edges of a triangle.

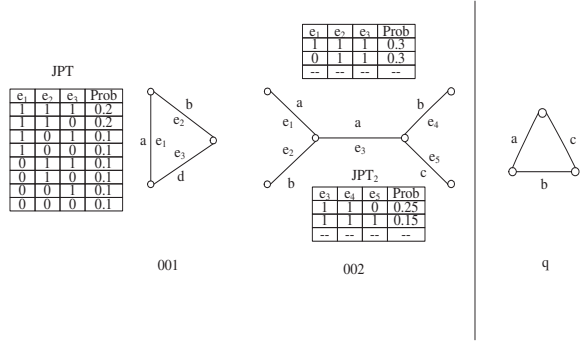

Figure 1: Probabilistic graph database & Query graph

## 1.1 Probabilistic Subgraph Matching

In this paper, we focus on *threshold-based probabilistic subgraph similarity matching* (T-PS) over a large set of probabilistic graphs. Specifically, let $D = \{g_1, g_2, ..., g_n\}$ be a set of probabilistic graphs where edges' existences are not independent, but are given explicitly by joint distributions, $q$ be a query graph, and $\epsilon$ be a probability threshold, a T-PS query retrieves all graphs $g \in D$ such that the *subgraph similarity probability* (SSP) between $q$ and $g$ is at least $\epsilon$. We will formally define SSP later (Def 9). We employ the *possible world semantics* [31, 11], which has been widely used for modeling probabilistic databases, to explain the meaning of returned results for subgraph similarity search. A possible world graph (PWG) of a probabilistic graph is a possible *instance* of the probabilistic graph. It contains all vertices and a subset of edges of the probabilistic graph, and it has a weight which is obtained by joining joint probability tables of all neighbor edges. Then, for a query graph $q$ and a probabilistic graph $g$, the probability that $q$ subgraph similarly matches $g$ is the summation of the weights of those PWGs, of $g$, to which $q$ is *subgraph similar*. If $q$ is subgraph similar to a PWG $g'$, $g'$ must contain a subgraph of $q$, say $q'$, such that the difference between $q$ and $q'$ must be less than the user specified error tolerance threshold $\delta$. In other words, $q$ is subgraph isomorphic to $g'$ after $q$ is relaxed with $\delta$ edges.

**Example** 1. *Consider graph 002 in Figure 1. $JPT_1$ and $JPT_2$ give joint distributions of neighbor edges $\{e_1, e_2, e_3\}$ and $\{e_3, e_4, e_5\}$ respectively. Figure 2 lists partial PWGs of probabilistic graph 002 and their weights. The weight of PWG (1) is obtained by joining $t_1$ of $JPT_1$ and $t_2$ of $JPT_2$, i.e., $Pr(e_1 = 1, e_2 = 1, e_3 = 1, e_4 = 1, e_5 = 0) = Pr(e_1 = 1, e_2 = 1, e_3 = 1) \times Pr(e_3 = 1, e_4 = 1, e_5 = 0) = 0.3 \times 0.25 = 0.075$. Suppose the distance threshold is 1. To decide if $q$ subgraph similarly matches probabilistic graph 002, we first find all of 002's PWGs that contain a subgraph whose difference between $q$ is less than 1. The results are PWGs (1), (2), (3) and (4), as shown in Figure 2, since we can delete edge a, b or c of q. Next, we add up the probabilities of these PWGs: $0.075 + 0.045 + 0.075 + 0.045 + ... = 0.45$. If the query specifies a probability threshold of 0.4, then graph 002 is returned since $0.45 > 0.4$.*

The above example gives a naive solution, to T-PS query processing, that needs to enumerate all PWGs of a probabilistic graph. This solution is very inefficient due to the exponential number of PWGs. Therefore, in this paper, we propose a *filter-and-verify* method to reduce the search space.

## 1.2 Overview of Our Approach

Given a set of probabilistic graphs $D = \{g_1, ..., g_n\}$ and a query graph $q$, our solution performs T-PS query in three steps, namely, structural pruning, probabilistic pruning, and verification.

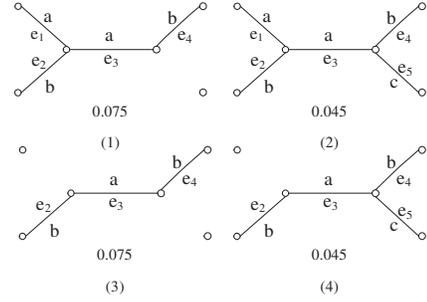

Figure 2: Partial possible world graphs of probabilistic graph 002

### Structural Pruning

The idea of structural pruning is straightforward. If we remove all the uncertainty in a probabilistic graph, and $q$ is still not subgraph similar to the resulting graph, then $q$ cannot subgraph similarly match the original probabilistic graph.

Formally, for $g \in D$, let $g^c$ denote the *corresponding deterministic graph* after we remove all the uncertain information from $g$. We have

**Theorem** 1. *If $q \not\subseteq_{sim} g^c$, $Pr(q \subseteq_{sim} g) = 0$.*

where $\subseteq_{sim}$ denotes subgraph similar relationship (Def 8), and $Pr(q \subseteq_{sim} g)$ denotes the subgraph similarity probability of $q$ to $g$.

Based on this observation, given $D$ and $q$, we can prune the database $D^c = \{g_1^c, ..., g_n^c\}$ using conventional deterministic graph similar matching methods. In this paper, we adopt the method in [38] to quickly compute results. [38] uses a multi-filter composition strategy to prune large number of graphs directly without performing pairwise similarity computation, which makes [38] more efficient compared to other graph similar search algorithms [15, 41]. Assume the result is $SC_q^c = \{g^c | q \subseteq_{sim} g^c, g^c \in D^c\}$. Then, its corresponding probabilistic graph set, $SC_q = \{g | g^c \in SC_q^c\}$, is the input for uncertain subgraph similar matching in the next step.

### Probabilistic Pruning

To further prune the results, we propose a *Probabilistic Matrix Index* (PMI) that will be introduced later, for probabilistic pruning. For a given set of probabilistic graphs $D$ and its corresponding set of deterministic graphs $D^c$, we create a feature set $F$ from $D^c$, where each feature is a deterministic graph, i.e., $F \subset D^c$. In PMI, for each $g \in SC_q$, we can locate a set

$$D_g = \{\langle LowerB(f_j), UpperB(f_j)\rangle | f_j \subseteq_{iso} g^c, 1 \leq j \leq |F|\}$$

where $LowerB(f)$ and $UpperB(f)$ are the lower and upper bounds of the subgraph isomorphism probability of $f$ to $g$ (Def 6), denoted by $Pr(f \subseteq_{iso} g)$. In this paper, $\subseteq_{iso}$ is used to denote subgraph-isomorphism. If $f$ is not subgraph isomorphic to $g^c$, we have $\langle 0 \rangle$.

In the probabilistic filtering, we first determine the remaining graphs after $q$ is relaxed with $\delta$ edges, where $\delta$ is the subgraph distance threshold. Suppose the remaining graphs are $\{rq_1, ...rq_i, ...rq_a\}$. For each $rq_i$, we compute two features $f_i^1$ and $f_i^2$ in $D_g$ such that $rq_i \supseteq_{iso} f_i^1$ and $rq_i \subseteq_{iso} f_i^2$. Let $Pr(q \subseteq_{sim} g)$ denote the subgraph similarity probability of $q$ to $g$ (Def 9). Then, we can calculate upper and lower bounds of $Pr(q \subseteq_{sim} g)$ based on the values of $UpperB(f_i^1)$ and $LowerB(f_i^2)$ for $1 \leq i \leq a$ respectively. If the upper bound of $Pr(q \subseteq_{sim} g)$ is smaller than probability threshold $\epsilon$, $g$ is pruned. If the lower bound of $Pr(q \subseteq_{sim} g)$ is not smaller than $\epsilon$, $g$ is in the final answers.

*Verification*

In this step, we calculate $Pr(q \subseteq_{sim} g)$ for query $q$ and candidate answer $g$, after probabilistic pruning, to make sure $g$ is really an answer, i.e. $Pr(q \subseteq_{sim} g) \geq \epsilon$.

## 1.3 Contributions and Paper Organization

The main idea (contribution) of our approach is to use the probabilistic checking of feature-based index to discriminate most graphs. To achieve this, several challenges need to be addressed.

*Challenge 1: Determine best bounds of $Pr(q \subseteq_{sim} g)$*

For each $rq_i$, we can find many $f_i^1$s and $f_i^2$s in PMI, thus, a large number of bounds of $Pr(q \subseteq_{sim} g)$ based on the combination of $UpperB(f_i^1)$ and $LowerB(f_i^2)$ for $1 \leq i \leq a$ can be computed. In this paper, we convert the problem of computing the best upper bound into a *set cover* problem. Our contribution is to develop an efficient randomized algorithm to obtain the best upper bound using integer quadratic programming, which is presented in Section 3.

*Challenge 2: Compute an effective $D_g$*

An effective $D_g$ should consist of tight $UpperB(f)$ and $LowerB(f)$ whose values can be computed efficiently. As we will show later that calculating $Pr(f \subseteq_{iso} g)$ is #P-complete, which increases the difficulty of computing an effective $D_f$. To address this challenge, we make a contribution to derive tight $UpperB(f)$ and $LowerB(f)$ by converting the problem of computing bounds into a maximum clique problem and propose an efficient solution by combining the properties of *probability conditional independence* and graph theory, which is discussed in Section 4.1.

*Challenge 3: Find the features that maximize pruning*

Frequent subgraphs (mined from $D^c$) are commonly used as features in graph matching. However, it would be impractical to index all of them. Our goal is to maximize the pruning capability with a small number of features. To achieve this goal, we consider two criteria in selecting features, the size of the feature and the number of disjoint embeddings that a feature has. A feature of small size and many embeddings is preferred. The details about feature selection are given in Section 4.2.

*Challenge 4: Compute SSP efficiently*

Though we are able to filter out a large number of probabilistic graphs, computing the exact SSP in the verification phase may still take quite some time and become the bottleneck in query processing. To address this issue, we develop an efficient sampling algorithm, based on the Monte Carlo theory, to estimate SSP with a high quality, which is presented in Section 5.

In addition, in Section 2, we formally define T-PS queries over probabilistic graphs and give the complexity of the problem in Section 2. We discuss the results of performance tests on real data sets in Section 6 and the related works in Section 7. We conclude our work in Section 8.

## 2. PROBLEM DEFINITION

In this section, we define some necessary concepts and show the complexity of our problem. Table 1 summarizes the notations used in this paper.

### 2.1 Problem Definition

**Definition** 1. (***Deterministic Graph***) *An undirected deterministic graph[2] $g^c$, is denoted as $(V, E, \Sigma, L)$, where $V$ is a set of vertices, $E$ is a set of edges, $\Sigma$ is a set of labels, and $L : V \cup E \rightarrow \Sigma$ is*

---
[2]In this paper, we consider undirected graphs, although it is straightforward to extend our methods to directed graphs.

| Symbol | Description |
|---|---|
| $D, SC_q, A_q$ | the probabilistic database set |
| $D^c, SC_q^c$ | the deterministic database |
| $g$ | the probabilistic graph |
| $\epsilon$ | the user-specified probability threshold |
| $\delta$ | the subgraph distance threshold |
| $f, q, g'\ g^c$ | the deterministic graph |
| $U = \{rq_1, .., rq_a\}$ | the remaining graph set after $q$ is relaxed with $\delta$ edges |
| $LowerB(f), UpperB(f)$ | the lower and upper bounds of SIP |
| $L_{sim}(q), U_{sim}q$ | the lower and upper bounds of SSP |
| $Brq_i, Bf_i, Bc_i$ | the Boolean variables of query, embedding and cut |
| $Ef, Ec$ | the set of embeddings and cuts |
| $IN$ | the set of disjoint embeddings |
| $F$ | the feature set |
| $Pr(x_{ne})$ | the joint probability distribution of neighbor edges |
| $Pr(q \subseteq_{iso} g)$ | the isomorphism between $q$ and $g$ |
| $Pr(q \subseteq_{sim} g)$ | the subgraph similarity probability between $q$ and $g$ |

Table 1: Notations

*a function that assigns labels to vertices and edges. A set of edges are* neighbor edges, *denoted by ne, if they are incident to the same vertex or the edges form a triangle in $g^c$.*

For example, consider graph 001 in Figure 1. Edges $e_1$, $e_2$ and $e_3$ are neighbor edges, since they form a triangle. Consider graph 002 in Figure 1. Edges $e_3$, $e_4$, and $e_5$ are also neighbor edges, since they are incident to the same vertex.

**Definition** 2. (***Probabilistic Graph***) *A probabilistic graph is defined as $g = (g^c, X_E)$, where $g^c$ is a deterministic graph, and $X_E$ is a binary random variable set indexed by $E$. An element $x_e \in X_E$ takes values 0 and 1, and denotes the existence possibility of edge $e$. A joint probability density function $Pr(x_{ne})$ is assigned to each neighbor edge set, where $x_{ne}$ denotes the assignments restricted to the random variables of a neighbor edge set, ne.*

A probabilistic graph has uncertain edges but deterministic vertices. The probability function $Pr(x_{ne})$ is given as a joint probability table of random variables of $ne$. For example, the probabilistic graph 002 in Figure 1 has 2 joint probability tables associated with 2 neighbor edge sets, respectively.

**Definition** 3. (***Possible World Graph***) *A possible world graph $g' = (V', E', \Sigma', L')$ is an instantiation of a probabilistic graph $g = ((V, E, \Sigma, L), X_E)$, where $V' = V$, $E' \subseteq E$, $\Sigma' \subseteq \Sigma$. We denote the instantiation from $g$ to $g'$ as $g \Rightarrow g'$.*

Both $g'$ and $g^c$ are deterministic graphs. But a probabilistic graph $g$ corresponds to one $g^c$ and multiple possible world graphs. We use $PWG(g)$ to denote the set of all possible world graphs derived from $g$. For example, Figure 2 lists 4 possible world graphs of the probabilistic graph 002 in Figure 1.

**Definition** 4. (***Conditional Independence***) *Let $X$, $Y$, and $Z$ be sets of random variables. $X$ is conditionally independent of $Y$ given $Z$ (denoted by $X \perp Y|Z$) in distribution $Pr$ if:*

$$Pr(X = x; Y = y|Z = z) = Pr(X = x|Z = z)$$
$$Pr(Y = y|Z = z)$$

*for all values $x \in dom(X)$, $y \in dom(Y)$ and $z \in dom(Z)$.*

Following real applications [9, 28, 18, 16], we assume that any two disjoint subsets of Boolean variables, $X_A$ and $X_B$ of $X_E$, are conditionally independent given a subset $X_C$ ($X_A \perp X_B | X_C$), if there is a path from a vertex in $A$ to a vertex in $B$ passing through $C$. Then, the probability of a possible world graph $g'$ is given by:

$$Pr(g \Rightarrow g') = \prod_{ne \in NS} Pr(x_{ne}) \quad (1)$$

where $NS$ is all the sets of neighbor edges of $g$.

For example, in probabilistic graph 002 of Figure 1, $\{e_1, e_2\} \perp \{e_4, e_5\} | e_3$. Clearly, for any possible world graph $g'$, we have $Pr(g \Rightarrow g') > 0$ and $\sum_{g' \in PWG(g)} Pr(g \Rightarrow g') = 1$, that is, each possible world graph has an existence probability, and the sum of these probabilities is 1.

**Definition 5.** *(Subgraph Isomorphism)* Given two deterministic graphs $g_1 = (V_1, E_1, \Sigma_1, L_1)$ and $g_2 = (V_2, E_2, \Sigma_2, L_2)$, we say $g_1$ is subgraph isomorphic to $g_2$ (denoted by $g_1 \subseteq_{iso} g_2$), if and only if there is an injective function $f : V_1 \to V_2$ such that:

- for any $(u, v) \in E_1$, there is an edge $(f(u), f(v)) \in E_2$;
- for any $u \in V_1$, $L_1(u) = L_2(f(u))$;
- for any $(u, v) \in E_1$, $L_1(u, v) = L_2(f(u), f(v))$.

*The subgraph $(V_3, E_3)$ of $g_2$ with $V_3 = \{f(v) | v \in V_1\}$ and $E_3 = \{(f(u), f(v)) | (u, v) \in E_1\}$ is called the embedding of $g_1$ in $g_2$.*

When $g_1$ is subgraph isomorphic to $g_2$, we also say that $g_1$ is a subgraph of $g_2$ and $g_2$ is a super-graph of $g_1$.

**Definition 6.** *(Subgraph Isomorphism Probability)* For a deterministic graph $f$ and a probabilistic graph $g$, we define their subgraph isomorphism probability (SIP) as,

$$Pr(f \subseteq_{iso} g) = \sum_{g' \in SUB(f,g)} Pr(g \Rightarrow g') \quad (2)$$

where $SUB(f, g)$ is $g$'s possible world graphs that are super-graphs of $f$, that is, $SUB(f, g) = \{g' \in PWG(g) | f \subseteq_{iso} g'\}$.

**Definition 7.** *(Maximum Common Subgraph-MCS)* Given two deterministic graphs $g_1$ and $g_2$, the maximum common subgraph of $g_1$ and $g_2$ is the largest subgraph of $g_2$ that is subgraph isomorphic to $g_1$, denoted by $mcs(g_1, g_2)$.

**Definition 8.** *(Subgraph Distance)* Given two deterministic graphs $g_1$ and $g_2$, the subgraph distance is, $dis(g_1, g_2) = |g_1| - |mcs(g_1, g_2)|$. Here, $|g_1|$ and $|mcs(g_1, g_2)|$ denote the number of edges in $g_1$ and $mcs(g_1, g_2)$, respectively. For a distance threshold $\delta$, if $dis(g_1, g_2) \leq \delta$, we call $g_1$ is subgraph similar to $g_2$.

Note that, in this definition, subgraph distance only depends on the edge set difference, which is consistent with pervious works on similarity search over deterministic graphs [38, 15, 30]. The operations on an edge consist of edge deletion, relabeling and insertion.

**Definition 9.** *(Subgraph Similarity Probability)* For a given query graph $q$, a probabilistic graph $g$[3] and a subgraph distance threshold $\delta$, we define their subgraph similarity probability as,

$$Pr(q \subseteq_{sim} g) = \sum_{g' \in SIM(q,g)} Pr(g \Rightarrow g') \quad (3)$$

where $SIM(q, g)$ is $g$'s possible world graphs that have subgraph distance to $q$ no larger than $\delta$, that is, $SIM(q, g) = \{g' \in PWG(g) | dis(q, g') \leq \delta\}$.

---
[3]Without loss of the generality, in this paper, we assume query graph is a connected deterministic graph, and probabilistic graph is connected.

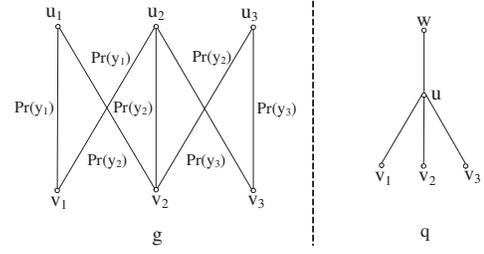

Figure 3: The probabilistic graph $g$ and query graph $q$ constructed for $(y_1 \wedge y_2) \vee (y_1 \wedge y_2 \wedge y_3) \vee (y_2 \wedge y_3)$.

| feature \ graph | 001 | 002 |
|---|---|---|
| $f_1$ | (0.55, 0.64) | (0.42, 0.5) |
| $f_2$ | (0.3, 0.48) | (0.26, 0.58) |
| $f_3$ | 0 | (0.08, 0.15) |

PMI

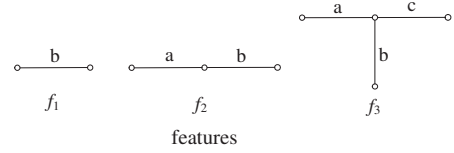

features

Figure 4: Probabilistic Matrix Index (PMI) & features of probabilistic graph database

**Problem Statement.** Given a set of probabilistic graphs $D = \{g_1, ..., g_n\}$, a query graph $q$, and a probability threshold $\epsilon$ ($0 < \epsilon \leq 1$), a subgraph similar query returns a set of probabilistic graphs $\{g | Pr(q \subseteq_{sim} g) \geq \epsilon, g \in D\}$.

## 2.2 Problem Complexity

From the problem statement, we know that in order to answer probabilistic subgraph similar queries efficiently, we need to calculate SSP (subgraph similarity probability) efficiently. We now show the time complexity of calculating SSP.

**Theorem 2.** *It is #P-complete to calculate the subgraph similarity probability.*

**Proof.** Due to space limit, we do not give the full proof and just highlight the major steps here. We consider a probabilistic graph whose edge probabilities are independent from each other. This probabilistic graph model is a special case of the probabilistic graph defined in Definition 2. We prove the theorem by reducing an arbitrary instance of the #P-complete DNF counting problem [13] to an instance of the problem of computing $Pr(q \subseteq_{sim} g)$ in polynomial time. Figure 3 illustrates an reduction for the DNF formula $F = (y_1 \wedge y_2) \vee (y_1 \wedge y_2 \wedge y_3) \vee (y_2 \wedge y_3)$. In the figure, the graph distance between $q$ and each possible world graph $g'$ is 1 (delete vertex $w$ from $q$). Each truth assignment to the variables in $F$ corresponds to a possible world graph $g'$ derived from $g$. The probability of each truth assignment equals to the probability of $g'$ that the truth assignment corresponds to. A truth assignment satisfies $F$ if and only if $g'$, the truth assignment corresponds to, is subgraph similar to $q$ (suppose graph distance is 1). Thus, $Pr(F)$ is equal to the probability, $Pr(q \subseteq_{sim} g)$. ∎

## 3. PROBABILISTIC PRUNING

As mentioned in Section 1.2, we first conduct structural pruning to remove probabilistic graphs that do not approximately contain the query graph $q$, and then we use probabilistic pruning techniques to further filter the remaining probabilistic graph set, named $SC_q$.

## 3.1 Pruning Conditions

We first introduce an index structure, *Probabilistic Matrix Index* (PMI), to facilitate probabilistic filtering. Each column of the matrix corresponds to a probabilistic graph in the database $D$, and each row corresponds to an indexed feature. Each entry records $\{LowerB(f), UpperB(f)\}$, where $UpperB(f)$ and $LowerB(f)$ are the upper and lower bounds of the subgraph isomorphism probability of $f$ to $g$, respectively.

**Example** 2. *Figure 4 shows the PMI of probabilistic graphs in Figure 1.*

Given a query $q$, a probabilistic graph $g$ and subgraph distance $\delta$, we generate a graph set, $U = \{rq_1, .., rq_a\}$, by relaxing $q$ with $\delta$ edge deletions or relabelings[4]. Here, we use the solution proposed in [38] to generate $\{rq_1, .., rq_a\}$. Suppose we have built the PMI. For each $g \in SC_q$, in PMI, we locate

$$D_g = \{\langle LowerB(f_j), UpperB(f_j)\rangle | f_j \subseteq_{iso} g^c, 1 \leq j \leq |F|\}$$

For each $rq_i$, we find two graph features in $D_g$, $\{f_i^1, f_i^2\}$, such that $rq_i \supseteq_{iso} f_i^1$ and $rq_i \subseteq_{iso} f_i^2$, where $1 \leq i \leq a$. Then we have probabilistic pruning conditions as follows.

**Pruning 1.**(*subgraph pruning*) Given a probability threshold $\epsilon$ and $D_g$, if $\sum_{i=1}^{a} UpperB(f_i^1) < \epsilon$, then $g$ can be safely pruned from $SC_q$.

**Pruning 2.**(*super graph pruning*) Given a probability threshold $\epsilon$ and $D_g$, if $\sum_{i=1}^{a} LowerB(f_i^2) - \sum_{1 \leq i,j \leq a} UpperB(f_i^2) UpperB(f_j^2) \geq \epsilon$, then $g$ is in the final answers, i.e., $g \in A_q$, where $A_q$ is the final answer set.

Before proving the correctness of the above two pruning conditions, we first introduce a lemma about $Pr(q \subseteq_{sim} g)$, which will be used for the proof. Let $Brq_i$ be a Boolean variable where $1 \leq i \leq a$, $Brq_i$ is true when $rq_i$ is subgraph isomorphic to $g^c$, and $Pr(Brq_i)$ is the probability that $Brq_i$ is true. We have

**Lemma** 1.
$$Pr(q \subseteq_{sim} g) = Pr(Brq_1 \vee ... \vee Brq_a). \qquad (4)$$

**Proof.** From Definition 9, we have
$$Pr(q \subseteq_{sim} g) = \sum_{g' \in SIM(q,g)} Pr(g \Rightarrow g') \qquad (5)$$

where $SMI(q,g)$ is a set of possible world graphs that have subgraph distance to $q$ no larger than $\delta$. Let $d$ be the subgraph distance between $q$ and $g^c$. We divide $SIM(q,g)$ into $\delta - d + 1$ subsets[5], $\{SM_0, ..., SM_{\delta-d}\}$, such that a possible world graph in $SM_i$ has subgraph distance $d + i$ with $q$. Thus, from Equation 5, we get

$$Pr(q \subseteq_{iso} g) = \sum_{g' \in SM_1 \cup ... \cup SM_{\delta-d}} Pr(g \Rightarrow g')$$
$$= \sum_{0 \leq j_1 \leq \delta-d} \sum_{g' \in SM_{j_1}} Pr(g \Rightarrow g') - \sum_{0 \leq j_1 < j_2 \leq \delta-d} \sum_{g' \in SM_{j_1} \cap SM_{j_2}} Pr(g \Rightarrow g') + \cdots + (-1)^i \sum_{0 \leq j_1 < ... < j_i \leq \delta-d} \sum_{g' \in SM_{j_1} \cap ... \cap SM_{j_i}} Pr(g \Rightarrow g')$$
$$+ \cdots + (-1)^{\delta-d} \sum_{g' \in SM_{j_1} \cap ... \cap SM_{j_{\delta-d}}} Pr(g \Rightarrow g'). \qquad (6)$$

---
[4]According to the subgraph similarity search, insertion does not change the query graph.
[5]For $g \in SC_q$, we have $d \leq \delta$, since the probabilistic graphs with $d > \delta$ have been filtered out in the deterministic pruning.

Let $L_i$, $0 \leq i \leq \delta - d$, be the graph set after $q$ is relaxed with $d + i$ edges, and $BL_i$ be a Boolean variable, when $BL_i$ is true, it indicates at least one graph in $L_i$ is a subgraph of $g^c$. Consider the $i$th item on the RHS in Equation 6, let $A$ be the set composed of all graphs in $i$ graph sets, and $B = BL_{j_1} \wedge ... \wedge BL_{j_i}$ be the corresponding Boolean variable of $A$. The set $g' \in SM_{j_1} \cap ... \cap SM_{j_i}$ contains all PWGs that have all graphs in $A$. Then, for the $i$th item, we get,

$$(-1)^i \sum_{0 \leq j_1 < ... < j_i \leq \delta-d} \sum_{g' \in SM_{j_1} \cap ... \cap SM_{j_i}} Pr(g \Rightarrow g')$$
$$= (-1)^i \sum_{0 \leq j_1 < ... < j_i \leq \delta-d} Pr(BL_{j_1} \wedge ... \wedge BL_{j_i}). \qquad (7)$$

Similarly, we can get the results for other items. By replacing the corresponding items with these results in Equation 6, we get

$$Pr(q \subseteq_{iso} g) = \sum_{0 \leq j_1 \leq \delta-d} Pr(BL_j) - \sum_{0 \leq j_1 < j_2 \leq \delta-d} Pr(BL_{j_1} \wedge BL_{j_2})$$
$$+ \cdots + (-1)^i \sum_{0 \leq j_1 < ... < j_i \leq \delta-d} Pr(BL_{j_1} \wedge ... \wedge BL_{j_i})$$
$$+ \cdots + (-1)^{\delta-d} Pr(BL_{j_1} \wedge ... \wedge BL_{j_{\delta-d}}). \qquad (8)$$

Based on the *Inclusion-Exclusion Principle* [26], the RHS of Equation 8 is $Pr(BL_0 \vee ... \vee BL_{\delta-d})$. Clearly, $BL_0 \subseteq ... \subseteq BL_{\delta-d}$, then
$$Pr(BL_0 \vee ... \vee BL_{\delta-d}) = Pr(BL_{\delta-d}) = Pr(Brq_1 \vee ... \vee Brq_a) \qquad \blacksquare$$

Lemma 1 gives a method to compute SSP. Intuitively, the probability of $q$ being subgraph similar to $g$ equals to the probability that at least one graph of the graph set $U = \{rq_1, ..., rq_a\}$ is a subgraph of $g$, where $U$ is remanning graph set after $q$ is relaxed with $\delta$ edges. With Lemma 1, we can formally prove the two pruning conditions.

**Theorem** 3. *Given a probability threshold $\epsilon$ and $D_g$, if $\sum_{i=1}^{a} UpperB(f_i^1) < \epsilon$, then $g$ can be safely pruned from $SC_q$.*

**Proof.** Since $rq_i \supseteq_{iso} f_i^1$, we have $Brq_1 \vee ... \vee Brq_a \subseteq Bf_1^1 \vee ... \vee Bf_a^1$, where $Bf_i^1$ is a Boolean variable denoting the probability of $f_i^1$ being a subgraph of $g$ for $1 \leq i \leq a$. Based on Lemma 1, we obtain

$$Pr(q \subseteq_{sim} g) = Pr(Brq_1 \vee ... \vee Brq_a)$$
$$\leq Pr(Bf_1^1 \vee ... \vee Bf_a^1)$$
$$\leq Pr(Bf_1^1) + ... + Pr(Bf_a^1)$$
$$\leq UpperB(f_1^1) + ... + UpperB(f_a^1) < \epsilon.$$

Then $g$ can be pruned. $\blacksquare$

**Theorem** 4. *Given a probability threshold $\epsilon$ and $D_g$, if $\sum_{i=1}^{a} LowerB(f_i^2) - \sum_{1 \leq i,j \leq a} UpperB(f_i^2) UpperB(f_j^2) \geq \epsilon$, then $g \in A_q$, where $A_q$ is the final answer set.*

**Proof.** Since $\vee_{i=1}^{a} Brq_i \supseteq \vee_{i=1}^{a} Bf_i^2$, we can show that

$$Pr(q \subseteq_{sim} g) = Pr(Brq_1 \vee ... \vee Brq_a)$$
$$\geq Pr(Bf_1^2 \vee ... \vee Bf_a^2)$$
$$\geq \sum_{i=1}^{a} Pr(Bf_i^2) - \sum_{1 \leq i,j \leq a} Pr(Bf_i^2) Pr(Bf_j^2)$$
$$\geq \sum_{i=1}^{a} LowerB(f_i^2) - \sum_{1 \leq i,j \leq a} UpperB(f_i^2) UpperB(f_j^2)$$
$$\geq \epsilon.$$

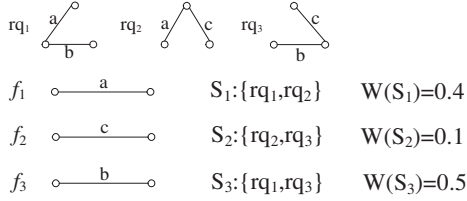

Figure 5: Obtain tightest $U_{sim}(q)$

Then $g \in A_q$. ∎

Note that the pruning process needs to address the traditional subgraph isomorphism problem ($rq \subseteq_{iso} f$ or $rq \supseteq_{iso} f$). In our work, we implement the state-of-the-art method VF2 [10].

## 3.2 Obtain Tightest Bounds of subgraph similarity probability

In pruning conditions, for each $rq_i$ ($1 \leq i \leq a$), we find only one pair feature $\{f_i^1, f_i^2\}$, among $|F|$ features, such that $rq_i \supseteq_{iso} f_i^1$ and $rq_i \subseteq_{iso} f_i^2$. Then we compute the upper bound, $U_{sim}(q) = \sum_{i=1}^{a} UpperB(f_i^1)$ and the lower bound $L_{sim}(q) = \sum_{i=1}^{a} LowerB(f_i^2) - \sum_{1 \leq i,j \leq a} UpperB(f_i^2) UpperB(f_j^2)$. However, there are many $f_i^1$s and $f_i^2$s satisfying conditions among $F$ features, therefore, we can compute a large number of $U_{sim}(q)$s and $L_{sim}(q)$-s. For each $rq_i$, if we find $x$ features meeting the needs among $|F|$ features, we can derive $x^a$ $U_{sim}(q)$s. Let $x = 10$ and $a = 10$, then there are $10^{10}$ upper bounds. The same holds for $L_{sim}(q)$. Clearly, it is unrealistic to determine the best bounds by enumerating all the possible ones, thus, in this section, we give efficient algorithms to obtain the tightest $U_{sim}(q)$ and $L_{sim}(q)$.

### 3.2.1 Obtain Tightest $U_{sim}(q)$

For each $f_j$ ($1 \leq j \leq |F|$) in PMI, we determine a graph set, $s_j$, that is a subset of $U = \{rq_1, ..., rq_a\}$, such that $rq_i \in s_j$ s.t. $rq_i \supseteq_{iso} f_j$. We also associate $s_j$ a weight, $UpperB(f_j)$. Then we obtain $|F|$ sets $\{s_1, .., s_{|F|}\}$ with each set having a weight $w(s_j) = UpperB(f_j)$ for $1 \leq j \leq |F|$. With this mapping, we transform the problem of computing tightest $U_{sim}(q)$ into a *weighted set cover* problem defined as follows.

**Definition** 10. (*Tightest $U_{sim}(q)$*) *Given a finite set $U = \{rq_1, ..., rq_a\}$ and a collection $S = \{s_1, .., s_j, .., s_{|F|}\}$ of subsets of $U$ with each $s_j$ attached a weight $w_{s_j}$, we want to compute a subset $C \subseteq S$ to minimize $\sum_{s_j \in C} w(s_j)$ s.t. $\bigcup_{s_j \in C} s_j = U$.*

It is well-known that the set cover problem is NP-complete [13], we use a greedy approach to approximate the tightest $U_{sim}(q)$. Algorithm 1 gives detailed steps. Assume the optimal value is OPT, the approximate value is within $OPT \cdot ln|U|$ [12].

---

**Algorithm 1** ObtainTightest$U_{sim}(q)(U, S)$

1: $A \leftarrow \phi, U_{sim}(q) = 0$;
2: **while** $A$ is not a cover of $U$ **do**
3:     for each $s \in S$, compute $\gamma(s) = \frac{w(s)}{|s - A|}$;
4:     choose an $s$ with minimal $\gamma(s)$;
5:     $A \leftarrow A \bigcup s$;
6:     $U_{sim}(q) + = w(s)$;
7: **end while**
8: **return** $U_{sim}(q)$;

---

**Example** 3. *In Figure 1, suppose we use $q$ to query probabilistic graph 002, and the subgraph distance is 1. The relaxed graph set of $q$ is $U = \{rq_1, rq_2, rq_3\}$ as shown in Figure 5. Given indexed features $\{f_1, f_2, f_3\}$, we first determine $s_1 = \{rq_1, rq_2\}$, $s_2 = \{rq_2, rq_3\}$ and $s_3 = \{rq_1, rq_3\}$. We use the $UpperB(f_j)$, $1 \leq j \leq 3$, as weight for three sets, and thus we have $w(s_1) = 0.4$, $w(s_2) = 0.1$ and $w(s_3) = 0.5$. Based on Definition 10, we obtain three $U_{sim}(q)$s, which are 0.4+0.1=0.5, 0.4+0.5=0.9 and 0.1+0.5=0.6. Finally the smallest (tightest) value, 0.5, is used as the upper bound, i.e., $U_{sim}(q) = 0.5$.*

### 3.2.2 Obtain Tightest $L_{sim}(q)$

For lower bound $L_{sim}(q)$, the larger (tighter) $L_{sim}(q)$ is, the better the probabilistic pruning power is. Here we formalize the problem of computing largest $L_{sim}(q)$ as an integer quadratic programming problem, and develop an efficient randomized algorithm to solve it.

For each $f_i$ ($1 \leq i \leq |F|$) in PMI, we determine a graph set, $s_i$, that is a subset of $U = \{rq_1, ..., rq_a\}$, such that $rq_j \in s_i$ s.t. $rq_j \subseteq_{iso} f_i$. We associate $s_i$ a pair weight of $\{LowerB(f_i), UpperB(f_i)\}$. Then we obtain $|F|$ sets $\{s_1, .., s_{|F|}\}$ with each set having a pair weight $\{w_L(s_i), w_U(s_i)\}$ for $1 \leq i \leq |F|$. Thus the problem of computing tightest $L_{sim}(q)$ can be formalized as follows.

**Definition** 11. (*Tightest $L_{sim}(q)$*) *Given a finite set $U = \{rq_1, ..., rq_a\}$ and a collection $S = \{s_1, ..., s_{|F|}\}$ of subsets of $U$ with each $s_i$ attached a pair weight $\{w_L(s_i), w_U(s_i)\}$, we want to compute a subset $C \subseteq \{s_1, ..., s_{|F|}\}$ to maximize*

$$\sum_{s_i \in C} w_L(s_i) - \sum_{s_i, s_j \in C} w_U(s_i) w_U(s_j)$$

*s.t. $\bigcup_{s_i \in C} s_i = U$.*

Associate an indicator variable, $x_{s_i}$, with each set $s_i \in S$, which takes value 1 if set $s_i$ is selected, 0 otherwise. Then we want to:

$$\begin{aligned} Maximize &\sum_{s_i \in C} x_{s_i} w_L(s_i) - \sum_{s_i, s_j \in C} x_{s_i} x_{s_j} w_U(s_i) w_U(s_j) \\ s.t. &\sum_{rq \in s_i} x_{s_i} \geq 1 \quad \forall rq \in U, \\ &x_s \in \{0, 1\}. \end{aligned} \quad (9)$$

Equation 9 is an integer quadratic programming which is a hard problem [13]. We *relax* $x_{s_i}$ to take values within $[0, 1]$, i.e., $x_{s_i} \in [0, 1]$. Then the equation becomes a standard quadratic programming (QP). Clearly, this QP is convex, and there is an efficient solution to solve the programming [23]. Since all feasible solutions for Equation 9 are also feasible solutions for the relaxed quadratic programming, the maximum value $QP(I)$ computed by the relaxed QP provides an upper bound for the value computed in Equation 9. Thus the value of $QP(I)$ can be used as the tightest lower bound. However, the proposed relaxation technique cannot give any theoretical guarantee on how tight $QP(I)$ is to Equation 9 [12].

Now following the relaxed QP, we propose a *randomized rounding* algorithm that yields an approximation bound for Equation 9. Algorithm 2 shows the detailed steps. According to Equation 9, it is not difficult to see that more elements in $U$ are covered, the tighter $L_{sim}(q)$ is. The following theorem states that the number of covered elements of $U$ has a theoretical guarantee.

**Theorem** 5. *When Algorithm 2 terminates, the probability that all elements are covered is at least $1 - \frac{1}{|U|}$.*

**Algorithm 2** ObtainTightest$L_{sim}(q)(U, S)$
1: $C \leftarrow \phi, L_{sim}(q) = 0$;
2: Let $x_s^*$ be an optimal solution to the relaxed QP;
3: **for** $k = 1$ to $2ln|U|$ **do**
4:    Pick each $s \in S$ independently with probability $x_s^*$;
5:    **if** $s$ is picked **then**
6:       $C \leftarrow s$;
7:       $L_{sim}(q) = L_{sim}(q) + w_L(s) - w_U(s) \sum_{l=1}^{|C|} w_U(s_l)$;
8:    **end if**
9: **end for**
10: **return** $L_{sim}(q)$;

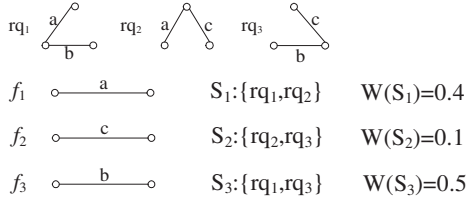

| | | |
|---|---|---|
| $f_1$    o———a———o | $S_1$:{rq$_1$,rq$_2$} | $W(S_1)=0.4$ |
| $f_2$    o———c———o | $S_2$:{rq$_2$,rq$_3$} | $W(S_2)=0.1$ |
| $f_3$    o———b———o | $S_3$:{rq$_1$,rq$_3$} | $W(S_3)=0.5$ |

Figure 6: Obtain tightest $L_{sim}(q)$

**Proof.** For an element $rq \in U$, the probability of $rq$ is not covered in an iteration is

$$\prod_{rq \in s}(1 - x_s^*) \leq \prod_{rq \in s} e^{-x_s^*} \leq e^{-\Sigma_{rq \in s} x_s^*} \leq \frac{1}{e}.$$

Then $rq$ is not covered at the end of the algorithm is at most $e^{-2log|U|} \leq \frac{1}{|U|^2}$. Thus, the probability that there is some $rq$ that is not covered is at most $|U| \cdot 1/|U|^2 = 1/|U|$. ∎

**Example** 4. *In Figure 1, suppose we use $q$ to query probabilistic graph 002, and the subgraph distance is 1. The relaxed graph set of $q$ is $U = \{rq_1, rq_2, rq_3\}$ shown in Figure 6. Given indexed features $\{f_1, f_2\}$, we first determine $s_1 = \{rq_1\}$ and $s_2 = \{rq_1, rq_2, rq_3\}$. Then we use $\{LowerB(f_i), UpperB(f_i)\}, 1 \leq i \leq 2$, as weights, and thus we have $\{w_L(s_1) = 0.28, w_U(s_1) = 0.36\}, \{w_L(s_2) = 0.08, w_U(s_2) = 0.15\}$. Based on Definition 11, we assign $L_{sim}(q) = 0.31$.*

## 4. PROBABILISTIC MATRIX INDEX

In this section, we discuss how to obtain tight $\{LowerB(f), UpperB(f)\}$ and generate features used in probabilistic matrix index (PMI).

### 4.1 Bounds of Subgraph Isomorphism Probability

#### 4.1.1 LowerB(f)

Let $Ef = \{f_1, .., f_{|Ef|}\}$ be the set of all embeddings[6] of feature $f$ in the deterministic graph $g^c$, $Bf_i$ be a Boolean variable for $1 \leq i \leq |Ef|$, which indicates whether $f_i$ exists in $g^c$ or not, and $Pr(Bf_i)$ be the probability of the embedding $f_i$ exists in $g$. Similar to Lemma 1, we have

$$Pr(f \subseteq_{iso} g) = Pr(Bf_1 \vee ... \vee Bf_{|Ef|}). \quad (10)$$

According to Theorem 2, it is not difficult to see that calculating the exact $Pr(f \subseteq_{iso} g)$ is NP-complete. Thus we rewrite Equation 10 as follows

---
[6]In this paper, we use the algorithm in [36] to compute embeddings of a $feature$ in $g^c$

$$\begin{aligned} Pr(f \subseteq_{iso} g) &= Pr(Bf_1 \vee ... \vee Bf_{|Ef|}) \\ &= 1 - Pr(\overline{Bf_1} \wedge ... \wedge \overline{Bf_{|Ef|}}) \\ &\geq 1 - Pr(\overline{Bf_1} \wedge ... \wedge \overline{Bf_{|IN|}} \mid \\ & \quad \overline{Bf_{|IN|+1}} \wedge ... \wedge \overline{Bf_{|Ef|}}). \end{aligned} \quad (11)$$

where $IN = \{Bf_1, ..., Bf_{|IN|}\} \subseteq Ef$.

Let the corresponding embeddings of $Bf_i$, $1 \leq i \leq |IN|$, do not have common parts (edges). Since $g^c$ is connected, these $|IN|$ Boolean variables are conditionally independent given any random variable of $g$. Then Equation 11 is written as

$$\begin{aligned} Pr(f \subseteq_{iso} g) &\geq 1 - Pr(\overline{Bf_1} \wedge ... \wedge \overline{Bf_{|IN|}} \mid \overline{Bf_{|IN|+1}} \wedge ... \wedge \overline{Bf_{|Ef|}}) \\ &= 1 - \prod_{i=1}^{|IN|}[1 - Pr(Bf_i \mid \overline{Bf_{|IN|+1}} \wedge ... \wedge \overline{Bf_{|Ef|}})]. \end{aligned} \quad (12)$$

For variables $Bf_x, Bf_y \in \{Bf_{|IN|+1}, ..., Bf_{|Ef|}\}$, we have

$$\begin{aligned} Pr(Bf_i|Bf_x \wedge Bf_y) &= \frac{Pr(Bf_i \wedge Bf_x \wedge Bf_y)}{Pr(Bf_x \wedge Bf_y)} \\ &= \frac{Pr(Bf_i \wedge Bf_x \wedge Bf_y)/Pr(Bf_y)}{Pr(Bf_x \wedge Bf_y)/Pr(Bf_y)} \\ &= \frac{Pr(Bf_i \wedge Bf_x|Bf_y)}{Pr(Bf_x|Bf_y)}. \end{aligned} \quad (13)$$

If $Bf_i$ and $Bf_x$ are conditionally independent given $Bf_y$, then

$$Pr(Bf_i \wedge Bf_x|Bf_y) = Pr(Bf_i|Bf_y)Pr(Bf_x|Bf_y). \quad (14)$$

By combining Equations 13 and 14, we obtain

$$Pr(Bf_i|Bf_x \wedge Bf_y) = Pr(Bf_i|Bf_y). \quad (15)$$

Based on this property, Equation 12 is reduced to

$$\begin{aligned} Pr(f \subseteq_{iso} g) &\geq 1 - \prod_{i=1}^{|IN|}[1 - Pr(Bf_i \mid \overline{Bf_{|IN|+1}} \wedge ... \wedge \overline{Bf_{|Ef|}})] \\ &= 1 - \prod_{i=1}^{|IN|}[1 - Pr(Bf_i \mid \overline{Bf_1} \wedge ... \wedge \overline{Bf_{|C|}})] \\ &= 1 - \prod_{i=1}^{|IN|}[1 - Pr(Bf_i|COR)] \end{aligned} \quad (16)$$

where $COR = \overline{Bf_1} \wedge ... \wedge \overline{Bf_{|C|}}$, and the corresponding embedding of $Bf_j \in C = \{Bf_1, ..., Bf_{|C|}\}$ overlaps with the corresponding embedding of $Bf_i$.

For a given $Bf_i$, $Pr(Bf_i|COR)$ is a constant, since the number of embeddings overlapping with $f_i$ in $g^c$ is constant. Now we obtain the lower bound of $Pr(f \subseteq_{iso} g)$ as

$$LowerB(f) = 1 - \prod_{i=1}^{|IN|}[1 - Pr(Bf_i|COR)], \quad (17)$$

which is only dependent on the selected $|IN|$ embeddings that do not have common parts with each other.

To compute $Pr(Bf_i|COR)$, a straightforward approach is the following. We first join all the joint probability tables (JPT), and meanwhile multiply joint probabilities of joining tuples in JPTs.

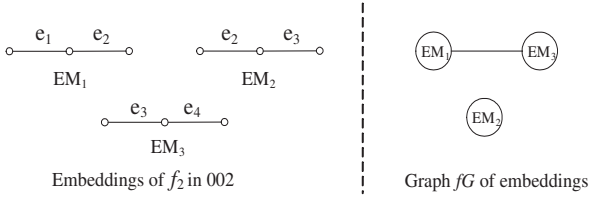

Figure 7: Embeddings & $fG$ of feature $f_2$ in probabilistic graph 002

Then, in the join result, we project on edge labels involved in $Bf_i$ and $COR$, and eliminate duplicates by summing up their existence probabilities. The summarization is the final result. However, this solution is clearly time inefficient for the sake of join, duplicate elimination, and probability multiplication.

In order to calculate $Pr(Bf_i|COR)$ efficiently, we use a sampling algorithm to estimate its value. Algorithm 3 shows the detailed steps. The main idea of the algorithm is as follows. We first sample a possible world $g'$. Then we check the condition, in Line 4, that is used to estimate $Pr(Bf_i \wedge COR)$, and the condition, in Line 7, that is used to estimate $Pr(COR)$. Finally we return $n_1/n_2$ which is an estimation of $Pr(Bf_i \wedge COR)/Pr(COR) = Pr(Bf_i|COR)$. The cycling number $m$ is set to $(4ln\frac{2}{\xi})/\tau^2$ ($0 < \xi < 1, \tau > 0$) used in Monte Carlo theory [26].

**Algorithm 3** Calculate $Pr(Bf_i|COR)$ $(g, Bf_i, COR)$
1: $n_1 = 0, n_2 = 0;$
2: **for** $i = 1$ to $m$ **do**
3:    Sample each neighbor edge set $ne$ of $g$ according to $Pr(x_{ne})$, and then obtain an instance $g'$;
4:    **if** $g'$ has embedding $f_i$ & no embeddings involved in $COR$ **then**
5:      $n_1 += 1;$
6:    **end if**
7:    **if** $g'$ has no embeddings involved in $COR$ **then**
8:      $n_2 += 1;$
9:    **end if**
10: **end for**
11: **return** $n_1/n_2;$

**Example 5.** *In Figure 4, consider $f_2$, a feature of probabilistic graph 002 shown in Figure 1. $f_2$ has three embeddings in 002, namely $EM1$, $EM2$ and $EM3$ as shown in Figure 7. In corresponding $Bf_i$s, $Bf_1$ and $Bf_3$ are conditionally independent given $Bf_2$. Then based on Equation 17, we have $LowerB(f) = 1 - [1 - Pr(Bf_1|\overline{Bf_2})][1 - Pr(Bf_3|\overline{Bf_2})] = 0.26$.*

As stated early, $LowerB(f)$ depends on embeddings that do not have common parts. However, among all $|Ef|$ embeddings, there are many groups which contain disjoint embeddings and leads to different lower bounds. We want to get a tight lower bound in order to increase the pruning power. Next, we introduce how to obtain tightest $LowerB(f)$.

**Obtain Tightest Lower Bound** We construct an undirected graph, $fG$, with each *node* representing an embedding $f_i$, $1 \leq i \leq |Ef|$, and a *link* connecting two *disjoint* embeddings (nodes). Note that, to avoid confusions, *nodes* and *links* are used for $fG$, while *vertices* and *edges* are for graphs. We also assign each node a weight, $-\ln[1 - Pr(Bf_i|COR)]$. In $fG$, a *clique* is a set of nodes such that any two nodes of the set are adjacent. We define the weight of a clique as the sum of node weights in the clique. Clearly, given a clique in $fG$ with weight $v$, $LowerB(f)$ is $1 - e^{-v}$. Thus, the larger the weight, the tighter (larger) the lower bound. To obtain a tight lower bound, we should find a clique whose weight is largest, which is exactly the *maximum weight clique* problem. Here we use the efficient solution in [7] to solve the maximum clique problem, and the algorithm returns the largest weight $z$. Therefore, we use $1 - e^{-z}$ as the tightest value for $LowerB(f)$.

**Example 6.** *Following Example 5, as shown in Figure 7, $EM1$ is disjoint with $EM3$. Based on the above discussion, we construct $fG$, for the three embeddings, shown in Figure 7. There are two maximum cliques namely, $\{EM1, EM3\}$ and $EM2$. According to Equation 17, the lower bounds derived from the 2 maximum cliques are 0.26 and 0.11 respectively. Therefore we select the larger (tighter) value 0.26 to be the lower bound of $f_2$ in 002.*

#### 4.1.2 $UpperB(f)$

Firstly, we define *Embedding Cut*: For a feature $f$, an embedding cut is a set of edges in $g^c$ whose removal will cause the absence of all $f$'s embeddings in $g^c$. An embedding cut is minimal if no proper subset of the embedding cut is an embedding cut. In this paper, we use minimal embedding cut.

Denote an embedding cut by $c$ and its corresponding Boolean variable (same as $Bf$) by $Bc$, where $Bc$ is true indicating that the embedding cut $c$ exists in $g^c$. Similar to Equation 10, it is not difficult to obtain,

$$Pr(f \subseteq_{iso} g) = 1 - Pr(Bc_1 \vee ... \vee Bc_{|Ec|}) \\ = Pr(\overline{Bc_1} \wedge ... \wedge \overline{Bc_{|Ec|}}) \quad (18)$$

where $Ec = \{c_1, ..., c_{|Ec|}\}$ is the set of all embedding cuts of $f$ in $g^c$. Equation 18 shows that the subgraph isomorphism probability of $f$ to $g$ equals the probability of all $f$'s embedding cuts disappearing in $g$.

Similar to the deduction from Equation 10 to 17 for $LowerB(f)$, we can rewrite Equation 18 as follows

$$\begin{aligned} Pr(f \subseteq_{iso} g) &= Pr(\overline{Bc_1} \wedge ... \wedge \overline{Bc_{|Ec|}}) \\ &\leq Pr(\overline{Bc_1} \wedge ... \wedge \overline{Bc_{|IN'|}}|\overline{Bc_{|IN'|+1}} \wedge ... \wedge \overline{Bc_{|Ec|}}) \\ &= \prod_{i=1}^{|IN'|}[1 - Pr(Bc_i|\overline{Bc_{|IN'|+1}} \wedge ... \wedge \overline{Bc_{|Ec|}})] \\ &= \prod_{i=1}^{|IN'|}[1 - Pr(Bc_i|\overline{Bc_1} \wedge ... \wedge \overline{Bc_{|D|}})] \\ &= \prod_{i=1}^{|IN'|}[1 - Pr(Bc_i|COM)] \end{aligned} \quad (19)$$

where $IN' = \{Bc_1, ..., Bc_{|IN'|}\}$ is a set of Boolean variables whose corresponding cuts are disjoint, $COM = \overline{Bc_1} \wedge ... \wedge \overline{Bc_{|D|}}$, and the corresponding cut of $Bc_j \in D = \{Bc_1, ..., Bc_{|D|}\}$ has common parts with the corresponding cut of $Bc_i$.

Finally we obtain the upper bound as

$$UpperB(f) = \prod_{i=1}^{|IN'|}[1 - Pr(Bc_i|COM)]. \quad (20)$$

The upper bound only relies on the picked embedding cut set in which any two cuts are disjoint.

The value of $Pr(Bc_i|COM)$ is estimated using Algorithm 3 by replacing embeddings with cuts. Similar to lower bound, computing tightest $UpperB(f)$ can be converted into a maximum weight clique problem. However, different from lower bound, each node of the constructed graph $fG$ represents a cut and has a weight of $-ln[1 - Pr(Bc_i|COM)]$ instead. Thus, for the maximum weight clique with weight $v$, the tightest value of $UpperB(f)$ is $e^{-v}$.

Now we discuss how to determine embedding cuts in $g^c$.

*Calculation of Embedding Cuts*

We build a connection between embedding cuts in $g^c$ and cuts for two vertices in a deterministic graph.

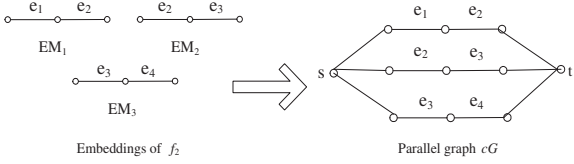

Figure 8: Transformation from embeddings of $f_2$ to parallel graph $cG$

Suppose $f$ has $|Ef|$ embeddings in $g^c$, and each embedding has $k$ edges. Assign $k$ labels, $\{e_1, ..., e_k\}$, for edges of each embedding (the order is random.). We create a corresponding *line* graph for each embedding by (1) creating $k + 1$ *isolated* nodes, and (2) connecting these $k + 1$ nodes to be a line by associating $k$ edges (with corresponding labels) of the embedding. Based on these line graphs, we construct a *parallel* graph, $cG$. The node set of $cG$ consists of all nodes of the $|Ef|$ line graphs and two new nodes, $s$ and $t$. The edge set of $cG$ consists of all edges (with labels) of the $|Ef|$ line graphs. In addition, one edge (without label) is placed between an end node of each line graph and $s$. Similarly, there is an edge between $t$ and the other end node of each line graph. As a result, $|Ef|$ embeddings are transformed into a deterministic graph $cG$.

Based on this transformation, we have

**Theorem** 6. *The embedding cut set of $g^c$ is also the cut set (without edges incident to $s$ and $t$) from $s$ to $t$ in $cG$.*

In this work, we determine embedding cuts using the method in [22].

**Example** 7. *Figure 8 shows the transformation for feature $f_2$ in graph 002 in Figure 1. In $cG$, we can find cuts $\{e_2, e_4\}$, $\{e_1, e_3, e_4\}$ and $\{e_2, e_3\}$ which are clearly the embedding cuts of $f_2$ in 002.*

### 4.2 Feature Generation

We would like to select frequent and discriminative features to construct probabilistic matrix index (PMI).

To achieve this, we consider $UpperB(f)$ given in Equation 20, since upper bound plays a most important role in the pruning capability. According to Equation 20, to get a tight upper bound, we need a large disjoint cut set and a large $Pr(Bc_i|COM)$. Suppose the cut set is $IN''$. Note that $|IN''| = |IN'|$, since a cut in $IN''$ has a corresponding Boolean variable $Bc_i$ in $IN'$. From the calculation of embedding cuts, it is not difficult to see that a large number of disjoint embeddings leads to a large $|IN''|$. Thus we would like a feature that has a large number of disjoint embeddings. Since $|COM|$ is small, a small size feature results in a large $Pr(Bc_i|COM)$. In summary, we should index a feature, which complies with following rules:

**Rule 1.** Select features that have a large number of disjoint embeddings.

**Rule 2.** Select small size features.

To achieve rule 1, we define the frequency of feature $f$ as $frq(f) = \frac{|\{g|f \subseteq_{iso} g^c, |IN|/|Ef| \geq \alpha, g \in D\}|}{|D|}$, where $\alpha$ is a threshold of the ratio of disjoint embeddings among all embeddings. Given a frequency threshold $\beta$, a feature $f$ is frequent iff $frq(f) \geq \beta$. Thus we would like to index a frequent feature. To achieve rule 2, we control a feature size used in Algorithm 4. To control feature number [37, 29], we also define the discriminative measure as: $dis(f) = \frac{|\cap \{D_{f'}|f' \subseteq_{iso} f\}|}{|D_f|}$, where $D_f$ is the list probabilistic graphs $g$ s.t. $f \subseteq_{iso} g^c$. Given a discriminative threshold $\gamma$, a feature $f$ is discriminative, iff $dis(f) > \gamma$. Thus we should also select a discriminative feature.

Based on the above discussion, we select frequent and discriminative features, which is implemented in Algorithm 4. In this algorithm, we first initial a feature set $F$ with single edge or vertex (line 1-4). Then we increase feature size (number of vertices) from 1, and pick out desirable features (line 6-9). $maxL$ is used to control the feature size, and guarantees picking out a small size feature satisfying rule 2. $frq(f)$ and $dis(f)$ are used to measure the frequency and discrimination of feature. The controlling parameters $\alpha$, $\beta$ and $\gamma$ guarantee picking out feature satisfying rule 1. The default values of the parameters are usually set to 0.1 [37, 38].

---
**Algorithm 4** FeatureSelection($D, \alpha, \beta, \gamma, maxL$)
---
1: $F \leftarrow \phi$;
2: Initial a feature set $F$ with single edge or vertex;
3: $D_f \leftarrow \{g|f \subseteq_{iso} g^c\}$;
4: $F \leftarrow F \cup \{f\}$;
5: **for** $i = 1$ to $maxL$ **do**
6:    **for** each feature $f$ with $i$ vertices **do**
7:       **if** $frq(f) \geq \beta$ & $dis(f) > \gamma$ **then**
8:          $D_f \leftarrow \{g|f \subseteq_{iso} g^c\}$;
9:          $F \leftarrow F \cup \{f\}$;
10:       **end if**
11:    **end for**
12: **end for**
13: **return** $F$;

## 5. VERIFICATION

In this section, we present the algorithms to compute subgraph similarity probability (SSP) of a candidate probabilistic graph $g$ to $q$.

Equation 4 is the formula to compute SSP. By simplifying this equation, we have

$$Pr(q \subseteq_{sim} g) = \sum_{i=1}^{a}(-1)^i \sum_{J \subseteq \{1,...,a\}, |J|=i} Pr(\wedge_{j=1}^{|J|} Brq_j). \quad (21)$$

Clearly, we need exponential number of steps to perform the exact calculation. Therefore, we develop an efficient sampling algorithm to estimate $Pr(q \subseteq_{sim} g)$.

By Equation 4, we know there are totally $a$ $Brq$s that are used to compute SSP. By Equation 10, we know $Brq = Bf_1 \vee ... \vee Bf_{|Ef|}$. Then, we have,

$$Pr(q \subseteq_{sim} g) = Pr(Bf_1 \vee ... \vee Bf_m) \quad (22)$$

where $m$ is the number of $Bf$s contained in these $a$ $Brq$s.

Assume $m$ $Bf$s have $x_1, ..., x_k$ Boolean variables for uncertain edges. Algorithm 5 gives detailed steps of the sampling algorithm. In this algorithm, we use *junction tree* algorithm to calculate $Pr(Bf_i)$ [17].

---
**Algorithm 5** Calculate $Pr(q \subseteq_{sim} g)$
---
1: $Cnt = 0$, $V = \sum_{i=1}^{m} Pr(Bf_i)$;
2: $N = (4ln2/\xi)/\tau^2$;
3: **for** 1 to $N$ **do**
4:    randomly choose $i \in \{1, ..., m\}$ with probability $Pr(Bf_i)/V$;
5:    randomly choose $x_1, .., x_k$ (according to probability $Pr(x_{ne})$) with $\{0, 1\}$ s.t. $Bf_i = 1$;
6:    **if** $Bf_1 = 0 \wedge ... \wedge Bf_{i-1} = 0$ **then**
7:       $Cnt = Cnt + 1$;
8:    **end if**
9: **end for**
10: **return** $Cnt/N$;

## 6. PERFORMANCE EVALUATION

In this section, we report the effectiveness and efficiency test results of our new proposed techniques. Our methods are implemented on a Windows XP machine with a Core 2 Duo CPU (2.8

GHz and 2.8 GHz) and 4GB main memory. Programs are compiled by Microsoft Visual C++ 6.0. In the experiments, we use a real probabilistic graph date set.

**Real Probabilistic Graph Dataset.** The real probabilistic graph dataset is obtained from the STRING database[7] that contains the protein-protein interaction (PPI) networks of organisms in the BioGRID database[8]. A PPI network is a probabilistic graph where vertices represent proteins, edges represent interactions between proteins, the labels of vertices are the COG functional annotations of proteins[9] provided by the STRING database, and the existence probabilities of edges are provided by the STRING database. We extract 5K probabilistic graphs from the database. The probabilistic graphs have an average number of 385 vertices and 612 edges. Each edge has an average value of 0.383 existence probability. According to [9], the neighbor PPIs (edges) are dominated by the strongest interactions of the neighbor PPIs. Thus, for each neighbor edge set $ne$, we set its probabilities as: $Pr(x_{ne}) = max_{1 \leq i \leq |ne|} Pr(x_i)$, where $x_i$ is a binary assignment to each edge in $ne$. Then, for each $ne$, we obtain $2^{|ne|}$ probabilities. We normalize those probabilities to construct the probability distribution, of $ne$, that is input into algorithms. Each query set $qi$ has 100 connected query graphs and query graphs in $qi$ are size-$i$ graphs (the edge number in each query is $i$), which are extracted from corresponding deterministic graphs of probabilistic graphs randomly, such as $q50$, $q100$, $q150$, $q200$ and $q250$. In scalability test, we randomly generate 2k, 4K, 6K, 8K and 10K data graphs.

The setting of experimental parameters is set as follows: the probability threshold is 0.3–0.7, and the default value is 0.5; the subgraph distance is 2–6, and the default value is 4; the query size is 50–250, and the default value is 150. In feature generation, the value of $maxL$ is 50–250, and the default value is 150; the values of $\{\alpha, \beta, \gamma\}$ are 0.05–0.25, and the default value is 0.15.

As introduced in Section 1.2, we implement the method in [38] to do structural pruning. This method is called *Structure* in experiments. In probabilistic pruning, the method using bounds of subgraph similarity probability is called *SSPBound*, and the approach using the best bounds is called *OPT-SSPBound*. To implement *SSPBound*, for each $rq_i$, we randomly find two features satisfying conditions in probabilistic matrix index (PMI). The method using bounds of subgraph isomorphism probability is called *SIPBound*, and the method using the tightest bound approach is called *OPT-SIPBound*. In verification, the sampling algorithm is called *SMP*, and the method given by Equation 21 is called *Exact*. Since there are no pervious works on the topic studied in this paper, we also compare the proposed algorithms with *Exact* that scans the probabilistic graph databases one by one. The complete proposed algorithm of this paper is called *PMI*. We report average results in following experiments.

In the first experiment, we demonstrate the efficiency of *SMP* against *Exact* in verification step. We first run structural and probabilistic filtering algorithms against the default dataset to create candidate sets. The candidate sets are then verified for calculating SSP using proposed algorithms. Figure 9(a) reports the result, from which we know *SMP* is efficient with average time less than 3 seconds, while the curve of *Exact* decreases in exponential. The approximation quality of *SMP* is measured by the *precision* and *recall* metrics with respect to query size shown in Figure 9(b). Precision is the percentage of true probabilistic graphs in the output probabilistic graphs. Recall is the percentage of returned probabilistic graphs in all true probabilistic graphs. The experimental results verify that *SMP* has a very high approximation quality with precision and recall both larger than 90%. We use *SMP* for verification in following experiments.

Figure 10 reports candidate sizes and pruning time of *SSPBound*, *OPT-SSPBound* and *Structure* with respect to probability thresholds. Recall that *SSPBound* and *OPT-SSPBound* are derived from upper and lower bounds of SIP. Here, we feed them with *OPT-SIPBound*. From the results, we know that the bars of *SSPBound* and *OPT-SSPBound* decrease with the increase of probability threshold, since larger thresholds can remove more false graphs with low confidences. As shown in Figure 10(a), the candidate size of *OPT-SSPBound* is very small (i.e., 15 on average), and is smaller than that of *SSPBound*, which indicates that our derived best bounds are tight enough to have a great pruning power. As shown in Figure 10(b), *OPT-SSPBound* has short pruning time (i.e., smaller than 1s on average) but takes more time than *SSPBound* due to more subgraph isomorphic tests during the calculation of *OPT-SSPBound*. Obviously, probabilities do not have impacts on *Structure*, and thus both bars of *Structure* hold constant.

Figure 11 shows candidate sizes and pruning time of *SIPBound*, *OPT-SIPBound* and *Structure* with respect to subgraph distance thresholds. To examine the two metrics, we feed *SIPBound* and *OPT-SIPBound* to *OPT-SSPBound*. From the results, we know that all bars increase with the increase of subgraph distance threshold, since larger thresholds lead to a large remaining graph set which is input into the proposed algorithms. Both *OPT-SIPBound* and *SIPBound* have a small number of candidate graphs, but *OPT-SIPBound* takes more time due to additional time for computing tightest bounds. From Figures 10(a) and 11(a), we believe that though *Structure* remains a large number of candidates, the probabilistic pruning algorithms can further remove most false graphs with efficient runtime. This observation verifies our algorithmic framework (i.e., structure pruning–probabilistic pruning– verification) is effective to process queries on a large probabilistic graph database.

Figure 12 examines the impact of parameters $\{maxL, \alpha, \beta, \gamma\}$ for feature generation. *Structure* holds constant in the 4 results, since the feature generation algorithm is used for probabilistic pruning. From Figure 12(a), we know the larger $maxL$ is, the more candidates *SSPBound* and *OPT-SSPBound* have. The reason is that the large $maxL$ generates large sized features, which leads to loose probabilistic bounds. From Figure 12(b), we see that all bars of probabilistic pruning first decrease and then increase, and reach lowest at the values 0.1 and 0.15 of $\alpha$. As shown in Figures 12(c) and 12(d), both bars of *OPT-SIPBound* decrease as the values of parameters increase, since either large $\beta$ or large $\gamma$ results in fewer features.

Figure 13 reports total query processing time with respect to different graph database sizes. *PMI* denotes the complete algorithm, that is, a combination of *Structure*, *OPT-SSPBound* (feed *OPT-SIPBound*) and *SMP*. From the result, we know *PMI* has quite efficient runtime and avoids the huge cost of computing SSP (#P-complete). PMI can process queries within 10 seconds on average. But the runtime of *Exact* grows in exponential, and has gone beyond 1000 seconds at the database size of 6k. The result of this experiment validates the designs of this paper.

Figure 14 examines the quality of query answers based on probability correlated and independent models. The query returns probabilistic graphs if the probabilistic graphs and the query (subgraph) belong to the same organism. We say the query and probabilistic graph belong to the same organism if the subgraph similarity probability is not less than the threshold. In fact the STRING database has given real organisms of the probabilistic graphs. Thus we can use the *precision* and *recall* to measure the query quality. Precision is the percentage of real probabilistic graphs in the returned probabilistic graphs. Recall is the percentage of returned real probabilistic graphs in all real probabilistic graphs. To determine query answers for the probability independent model, we multiply probabilities of edges in each neighbor edge set to obtain joint probability tables (JPT). Based on the JPTs, we use $PMI$ to determine query answers for the probability independent model. Each time, we randomly generate 100 queries and report average results. In the examination, $COR$ and $IND$ denote the probability correlated and probability independent models respectively. In the figure, *precision* and *recall* go down as probability threshold is larger, since large thresholds make query and graphs more difficult to be categorized into the same organism. We also know that the probability correlated model has much higher precision and recall than the probability independent model. The probability correlated model has average precision and recall both larger than 85%, while the probability independent model has values smaller than 60% at threshold larger than 0.6. The result indicates that our proposed model behaves more accurate biologic features than the probability independent model.

---

[7] http://string-db.org
[8] http://thebiogrid.org
[9] http://www.ncbi.nih.gov/COG

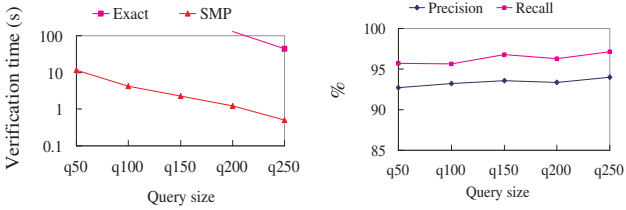

(a) Runtime  (b) Query quality

Figure 9: Scalability to query size.

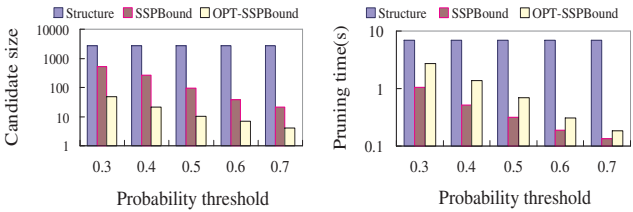

(a) Candidate size  (b) Runtime

Figure 10: Scalability to probability threshold.

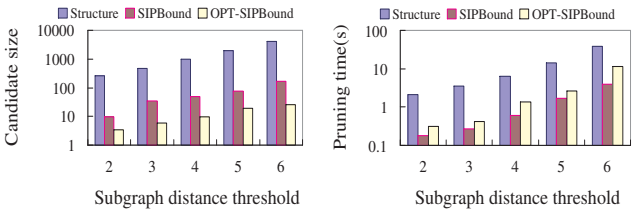

(a) Candidate size  (b) Runtime

Figure 11: Scalability to subgraph distance threshold.

## 7. RELATED WORK

In this paper, we study similarity search over uncertain graphs, which is related to uncertain and graph data management. Readers who are interested in general uncertain and graph data management please refer to [3] and [4] respectively.

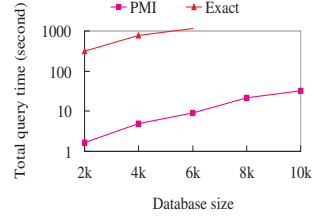

Figure 13: Total query processing time.

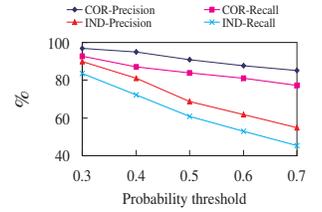

Figure 14: Query quality comparison (COR V.S. IND).

The topic most related to our work is similarity search in deterministic graphs. Yan et al. [38] proposed to process subgraph similarity queries based on frequent graph features. They used a filtering-verification paradigm to process queries. He et al [15] employed an R-tree like index structure, organizing graphs hierarchically in a tree, to support $k$-NN search to the query graph. Jiang et al [19] encoded graphs into strings and converted graph similarity search into string matching. Williams et al [35] aimed to find graphs with the minimum number of miss-matchings of vertex and edge labels bounded by a given threshold. Zeng et al [41] proposed tight bounds of graph edit-distance to filter out false graphs in similarity search, based on which, Wang et al [34] developed an indexing strategy to speed up query. Shang et al [30] studied super-graph similarity search, and proposes top-down and bottom-up index construction strategy to optimize the performance of query processing. Recently, Sun et al [32] proposed a subgraph matching algorithm on distributed in-memory graphs without using structured index.

Another related topic is querying uncertain graphs. Potamias et al [27] studied $k$-nearest neighbor queries ($k$-NN) over uncertain graphs, i.e., computing the $k$ closest nodes to a query node. They proposed sampling algorithms to answer the #P-complete k-NN queries. Zou et al [42, 43] studied frequent subgraph mining on uncertain graph data under the probability and expectation semantics respectively. Yuan et al [40] proposed graph feature-based framework to conduct uncertain subgraph graph query. In another work, Yuan et al [39] and Jin et al [21] studed shortest path query and distance-constraint reachability query in a single uncertain graph. The above works define uncertain graph models with independent edge distributions and do not consider edge correlations.

## 8. CONCLUSION

This is the first work to answer the subgraph similarity query on a large probabilistic graphs with correlation on edge probability distributions. Though it is an NP-hard problem, we employ the filter-and-verify methodology to answer the query efficiently. During the filtering phase, we propose a probabilistic matrix (PMI) index with tight upper and lower bounds of subgraph isomorphism probability. Based on PMI, we derive upper and lower bounds of subgraph similarity probability, and we compute best bounds by developing deterministic and randomized optimization algorithms. We also propose selective strategies for picking powerful subgraph features. Therefore we are able to filter out large number of probabilistic graphs without calculating the subgraph similar probabilities. During verification, we use the Monte Carlo theory to fast validate final answers with a high quality. Finally, we confirm our designs through an extensive experimental study.

## 9. ACKNOWLEDGMENT

Ye Yuan and Guoren Wang are supported by the NSFC (Grant No. 61025007, 60933001 and 61100024), National Basic Research

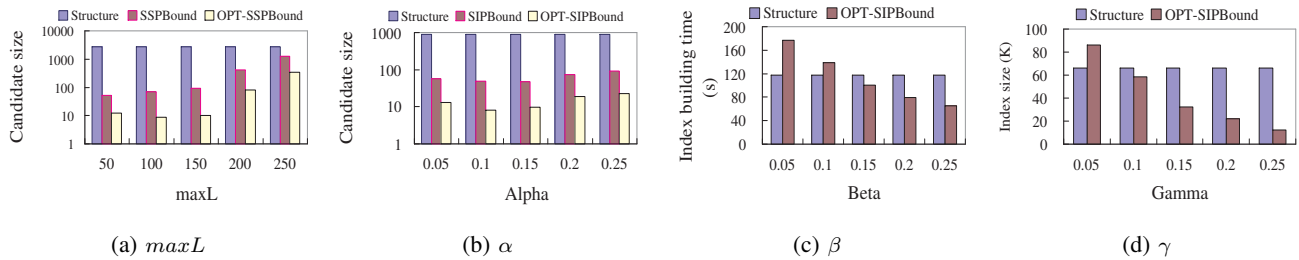

Figure 12: Impact of parameters for feature generation.

(a) $maxL$  (b) $\alpha$  (c) $\beta$  (d) $\gamma$

Program of China (973, Grant No. 2011CB302200-G), and the Fundamental Research Funds for the Central Universities (Grant No. N110404011). Lei Chen is supported by the Microsoft Research Asia Theme-based Grant under project MRA11EG05.